\documentstyle[12pt]{article}

\catcode`\@=11
\long\def\@makefntext#1{
\protect\noindent \hbox to 3.2pt {\hskip-.9pt
$^{{\ninerm\@thefnmark}}$\hfil}#1\hfill}                

 \def\@makefnmark{\hbox to 0pt{$^{\@thefnmark}$\hss}}  

\def\ps@myheadings{\let\@mkboth\@gobbletwo
\def\@oddhead{\hbox{}
\rightmark\hfil\ninerm\thepage}
\def\@oddfoot{}\def\@evenhead{\ninerm\thepage\hfil
\leftmark\hbox{}}\def\@evenfoot{}
\def\sectionmark##1{}\def\subsectionmark##1{}}

\newcounter{sectionc}\newcounter{subsectionc}\newcounter{subsubsectionc}
\renewcommand{\section}[1] {\vspace{0.6cm}\addtocounter{sectionc}{1}
\setcounter{subsectionc}{0}\setcounter{subsubsectionc}{0}\noindent
        {\bf\thesectionc. #1}\par\vspace{0.4cm}}
\renewcommand{\subsection}[1] {\vspace{0.6cm}\addtocounter{subsectionc}{1}
        \setcounter{subsubsectionc}{0}\noindent
        {\it\thesectionc.\thesubsectionc. #1}\par\vspace{0.4cm}}
\renewcommand{\subsubsection}[1]
{\vspace{0.6cm}\addtocounter{subsubsectionc}{1}
        \noindent {\rm\thesectionc.\thesubsectionc.\thesubsubsectionc.
        #1}\par\vspace{0.4cm}}

\newcounter{appendixc}
\newcounter{subappendixc}[appendixc]
\newcounter{subsubappendixc}[subappendixc]

\renewcommand{\appendix}[1] {\vspace{0.6cm}
        \refstepcounter{appendixc}
        \setcounter{figure}{0}
        \setcounter{table}{0}
        \setcounter{equation}{0}
        \renewcommand{\thefigure}{\Alph{appendixc}.\arabic{figure}}
        \renewcommand{\thetable}{\Alph{appendixc}.\arabic{table}}
        \renewcommand{\theappendixc}{\Alph{appendixc}}
        \renewcommand{\theequation}{\Alph{appendixc}.\arabic{equation}}
        \noindent{\bf Appendix \theappendixc #1}\par\vspace{0.4cm}}

\def\abstracts#1{{
        \centering{\begin{minipage}{30pc}\tenrm\baselineskip=12pt\noindent
        \centerline{\tenrm ABSTRACT}\vspace{0.3cm}
        \parindent=0pt #1
        \end{minipage}}\par}}

\renewenvironment{thebibliography}[1]
        {\begin{list}{\arabic{enumi}.}
        {\usecounter{enumi}\setlength{\parsep}{0pt}
\setlength{\leftmargin 1.25cm}{\rightmargin 0pt}
         \setlength{\itemsep}{0pt} \settowidth
        {\labelwidth}{#1.}\sloppy}}{\end{list}}

\newcounter{itemlistc}
\newcounter{romanlistc}
\newcounter{alphlistc}
\newcounter{arabiclistc}

\newcommand{\fcaption}[1]{
        \refstepcounter{figure}
        \setbox\@tempboxa = \hbox{\tenrm Fig.~\thefigure. #1}
        \ifdim \wd\@tempboxa > 6in
           {\begin{center}
        \parbox{6in}{\tenrm\baselineskip=12pt Fig.~\thefigure. #1}
            \end{center}}
        \else
             {\begin{center}
             {\tenrm Fig.~\thefigure. #1}
              \end{center}}
        \fi}

\newcommand{\tcaption}[1]{
        \refstepcounter{table}
        \setbox\@tempboxa = \hbox{\tenrm Table~\thetable. #1}
        \ifdim \wd\@tempboxa > 6in
           {\begin{center}
        \parbox{6in}{\tenrm\baselineskip=12pt Table~\thetable. #1}
            \end{center}}
        \else
             {\begin{center}
             {\tenrm Table~\thetable. #1}
              \end{center}}
        \fi}

\def\@citex[#1]#2{\if@filesw\immediate\write\@auxout
        {\string\citation{#2}}\fi
\def\@citea{}\@cite{\@for\@citeb:=#2\do
        {\@citea\def\@citea{,}\@ifundefined
        {b@\@citeb}{{\bf ?}\@warning
        {Citation `\@citeb' on page \thepage \space undefined}}
        {\csname b@\@citeb\endcsname}}}{#1}}

\newif\if@cghi
\def\cite{\@cghitrue\@ifnextchar [{\@tempswatrue
        \@citex}{\@tempswafalse\@citex[]}}
\def\citelow{\@cghifalse\@ifnextchar [{\@tempswatrue
        \@citex}{\@tempswafalse\@citex[]}}
\def\@cite#1#2{{$\null^{#1}$\if@tempswa\typeout
        {IJCGA warning: optional citation argument
        ignored: `#2'} \fi}}

\def\fnt#1#2{\footnotetext{\kern-.3em
        {$^{\mbox{\sevenrm #1}}$}{#2}}}

 1
 1
 1

\font\tenbf=cmbx10
\font\tenrm=cmr10
\font\tenit=cmti10

\font\ninerm=cmr9

\newcommand{\be}{\begin{equation}}
\newcommand{\ee}{\end{equation}}

\newcommand{\bea}{\begin{eqnarray}}
\newcommand{\eea}{\end{eqnarray}}
\newcommand{\ba}{\begin{array}}
\newcommand{\ea}{\end{array}}
\newcommand{\lsim}{{\;\raise0.3ex\hbox{$<$\kern-0.75em\raise-1.1ex\hbox{$\sim$}}
\;}}
\newcommand{\gsim}{{\;\raise0.3ex\hbox{$>$\kern-0.75em\raise-1.1ex\hbox{$\sim$}}
\;}}

\newcommand{\CR}{\nonumber \\}

\newcommand{\de}{\delta}
\newcommand{\DE}{\Delta}
\newcommand{\D}{\delta}

\newcommand{\ssu}{$SU(2)_L\times SU(2)_R\times U(1)_{B-L}\,$}

\newcommand{\matr}{\left( \begin{array}}
\newcommand{\ematr}{\end{array} \right)}
\newcommand{\g}{\gamma}

\newcommand{\lr}{{left-right symmetric model$\:$}}

\begin{document}

\mbox{}\vspace*{-1cm}\hspace*{8cm}\makebox[7cm][r]{\large  HU-SEFT R 1994-19}
\vspace*{2.0cm}
 \hspace*{10.9cm} \makebox[4cm]{(hep-ph/yymmxxx)}

\centerline{\tenbf PHENOMENOLOGICAL IMPLICATIONS OF SUPERSYMMETRY}
\baselineskip=16pt
\centerline{\tenbf IN LEFT-RIGHT ELECTROWEAK MODEL\footnote{Talk given
by Martti Raidal in "The First Arctic Workshop on Future Physics and
Accelerators,"  Saariselk\"a, Finland August 21-26, 1994.}}
\vspace{0.8cm}
\centerline{\tenrm K. HUITU, M. RAIDAL}
\baselineskip=13pt
\centerline{\tenit Research Institute for High Energy Physics}
\centerline{\tenit P.O. Box 9, FIN-00014 University of Helsinki, Finland }
\vspace{0.3cm}
\centerline{\tenrm and}
\vspace{0.3cm}
\centerline{\tenrm J. MAALAMPI}
\baselineskip=13pt
\centerline{\tenit Department of Theoretical Physics }
\centerline{\tenit P.O. Box 9, FIN-00014 University of Helsinki, Finland }
\vspace{0.9cm}
\abstracts{The basics of a supersymmetric \ssu model are reviewed.
The production and subsequent decays of the doubly charged triplet
higgsino $ \tilde\Delta^{\pm\pm}$ in the Next Linear Collider are
discussed. The slepton pair production in the framework of
this model is also analysed.
}

\vfil
\rm\baselineskip=14pt

\section{Introduction}

Despite  of the success of the Standard Model (SM)
there are still unsolved problems in particle physics which motivate
searches for  more fundamental theories. One of the
most appealing extensions of the SM is the \lr
based on the gauge group \ssu \cite{pati}.
Among many attractive features of this model is its capability to
explain  the lightness of the ordinary neutrinos via
so called see-saw mechanism \cite{seesaw}
which occurs naturally due to the dynamical
treatment of both left and right handed fields.
Indeed, the anomalies measured in
the solar \cite{sun} and atmospheric \cite{atmos}
 neutrino fluxes
as well as the COBE observation \cite{cobe}
of the existence of the hot component
of dark matter
seem to indicate that neutrinos could have a small non-vanishing mass.
Another peculiar feature of the model is the existence
of  lepton number violating interactions, partly due to the massive Majorana
neutrinos and partly because of doubly charged components of triplet
Higgs fields which carry lepton number two.

On the other hand the \lr similarly to the SM suffers from the hierarchy
problem: the masses of the Higgs scalars diverge quadratically.
As in the SM, supersymmetry can be used to cure this problem.

Here we shall consider
 a minimal susy left-right model where the number of Higgs fields
is the smallest possible \cite{meie}. In particular we shall discuss
 how  one can test the model in future  collider experiments.
Since the doubly charged triplet higgsinos give the most distinctive
experimental signature we shall study the triplet higgsino production
in possible $ e^-e^-,$ $ e^+e^-,$ $ e^-\g$ and $ \g\g$ options of the
Next Linear Collider (NLC) \cite{wiik}. If the doubly charged
higgsino is too heavy to be produced in the NLC we shall study
how its extra contribution as a virtual intermediate state would
affect the selectron production in $e^+e^-$ collisions.

\section{A supersymmetric left-right model}

The minimal set of Higgs fields in the non-susy left-right model
consists of a bidoublet $ \phi_u$
and a $ SU(2)_R $ triplet $ \DE$.
In supersymmetrization, the cancellation of chiral anomalies among the
fermionic partners of the triplet Higgs fields $\DE$ requires
 introduction of the second triplet $\de$ with opposite $U(1)_{B-L}$
quantum number. Due to the conservation of the $B-L$ symmetry $\de$
does not couple to leptons and quarks. In order to avoid a trivial
Kobayashi-Maskawa matrix for quarks, also another bidoublet $ \phi_d$
 should be
added to the model. This is because supersymmetry forbids a Yukawa
coupling where the bidoublet appears as a conjugate.

We have chosen the vacuum expectation values for the Higgses, which
break the  $SU(2)_L\times SU(2)_R\times U(1)_{B-L}$ into the
$U(1)_{em}$,  to be as follows
 \be <\phi_u >=\left( {\begin{array}{cc} \kappa_u & 0\\ 0 & 0
\end{array}}  \right), \: <\phi_d >=\left( {\begin{array}{cc} 0 & 0\\
0 & \kappa_d \end{array}}  \right),\:   <\Delta>=\left(
{\begin{array}{cc} 0 & 0\\ v & 0 \end{array}} \right)
 ,\:   <\delta >\equiv 0. \label{vevs} \ee
\noindent Here  $\kappa_{u,d}$ are of the order of the  electroweak
scale $10^2$ GeV. The vev  $v$ of the triplet Higgs has to be much
larger in order to have the masses of the new gauge bosons $W_2$ and
$Z_2$ sufficiently high \cite{tevatron}.
With the choice (\ref{vevs}) of the vev's
the charged gauge bosons do not mix and $W_L$ corresponds to the
observed  particle.

We assume the superpotential  to have the following form:
 \bea W & = & h_u^Q \widehat Q_L^{cT} \widehat \phi_u  \widehat Q_R  +
h_d^Q \widehat Q_L^{cT} \widehat \phi_d  \widehat Q_R \nonumber \\
&&+h_u^L \widehat L_L^{cT} \widehat \phi_u  \widehat L_R  +h_d^L
\widehat L_L^{cT} \widehat \phi_d  \widehat L_R    +h_\DE \widehat
L_R^{T} i\tau_2 \widehat \DE  \widehat L_R \nonumber\\ && + \mu_1
{\rm Tr} (\tau_2 \widehat \phi_u^T \tau_2 \widehat \phi_d )  +\mu_2
{\rm Tr} (\widehat \DE \widehat \D ) .\label{pot} \eea

 \noindent Here $\widehat Q_{L(R)}$ stands for the
doublet of left(right)-handed quark superfields, $\widehat L_{L(R)}$
stands for the doublet of left(right)-handed lepton superfields,
$\widehat \phi_u$ and $\widehat \phi_d$ are the two bidoublet Higgs
superfields, and $\widehat \DE$ and $ \widehat \D$ the two triplet
Higgs superfields.

In the superpotential (\ref{pot}) the $R$-parity,
$R=(-1)^{3(B-L)+2S}$, is preserved. This ensures that the susy
partners with $R=-1$ are produced in pairs and  that the lightest
supersymmetric particle (LSP) is stable.
In
order  to preserve the naturalness of the theory the
 supersymmetric mass parameters $\mu_i $  should be
close to the  scale of the soft supersymmetry breaking\cite{haber}.
 We assume here that the parameters
$|\mu_i|$ are of the order of the weak scale.

We are especially interested in the doubly charged
fermions occurring in the Higgs triplet superfields. Their mass
matrix is particularly simple, since doubly charged higgsinos do not
mix with gauginos.  The triplet higgsinos, like the triplet
 Higgses, carry lepton number two and therefore
 the final state of their  decay must also have
even lepton number in the case of $R$-parity conservation.

 There are five charginos
$ \psi^{\pm}_j$
and nine neutralinos
$\psi^{0}_i$
in this model.
The physical particles $ \tilde\chi^{\pm}_i$
and $\tilde\chi^{0}_i $ are  found by  diagonalization of the mass
Lagrangian:
\be \tilde\chi^{\pm}_i=\sum_{j}C_{ij}^{\pm}\psi_j^{\pm},
\label{charginos} \ee
\be \tilde\chi^{0}_i=\sum_jN_{ij}\psi^0_j. \label{chargino} \ee

 The
masses of the particles depend on the following parameters: the soft gaugino
masses, the supersymmetric Higgs masses $\mu_1$ and
$\mu_2$, the vacuum expectation values $\kappa_u$, $\kappa_d$, and
$v$, and the gauge coupling constants.
 We have calculated numerically the composition of neutralinos and
charginos for  different values of the parameters. The neutralinos
are Majorana particles, whereas the charginos combine together  to
form Dirac fermions. For our numerical calculations we have always
taken $g_R=g_L=0.65,$ $h_{\DE}=0.3,$
$\mu_1=200\,$GeV and  $m_{W_R}=500\,$GeV. The parameter
$\mu_2$ is equal to the doubly charged higgsino mass $M_\DE$.
In the following
we shall consider two different sets of the model parameters.
In the first case the soft gaugino masses are taken to be
1 TeV (LRM I) and in the second case 200 GeV (LRM II).
For the larger soft gaugino masses  the neutralinos are predominantly
higgsinos whereas for the smaller soft masses they are mainly gauginos.
\begin{figure}

\vspace{6cm}


\caption{
 Branching ratios of (a) the left slepton  and
(b)  the right slepton as  functions of the slepton
mass for \(\mu_2=\)300GeV.}
\end{figure}

The mixing of the left and right selectrons
is assumed to be negligible  and their masses
$m_{\tilde e L}$ and
$m_{\tilde e R}$ are taken to be equal.

\section{Tests of the model in the NLC}

 The allowed decay modes of the triplet higgsinos are

\bea \tilde \Delta^{++}  & \rightarrow &  \Delta^{++}\, \lambda^{0}
,\:\:
 \Delta^{+}\, \lambda^{+},\CR &&
 \tilde \Delta^{+} \, W_2^+ , \:\: \tilde l^+ l^+ . \eea
 \noindent In large regions of the parameter space, the
kinematically favoured decay mode is  $\tilde \Delta^{++}
\rightarrow  \tilde l^+ l^+ $.   As the   masses  of
 $\Delta$ and $ W_2$ are of the order of the SU(2)$_R$ breaking
scale $v$ \cite{Gunion}, the first three decay channels are kinematically
forbidden in our case of the relatively light triplet higgsinos.
In the following we
shall assume that $\tilde\Delta^{++}$
 decay in 100\% into the $\tilde ll$ final
state.

The charged sleptons $\tilde l$ can decay as follows:
 \be \tilde l^+\to l^+
+ \tilde\chi_i^0, \label{decay1} \ee \be \tilde l^+\to \nu +
\tilde\chi_i^+, \label{decay2} \ee \be \tilde l^+\to W^+ + \tilde\nu.
\label{decay3} \ee The decay mode (\ref{decay3}) is kinematically
disfavoured and we do not consider it.
The decay of the the
right-slepton into the neutrino channel will in general be
kinematically disfavoured because of the heaviness
of the right-handed neutrino. In Fig. 1  the branching ratios of
the different channels are plotted as the function of the
left-slepton and right-slepton masses.
For the left-slepton decay the channel (\ref{decay2})
becomes dominant immediately when the slepton mass exceeds the
 mass of the lightest chargino. The chargino has several decay
channels, e.g. into a lepton-slepton pair, a W-chargino pair, and a
quark-squark pair.
\begin{figure}
\vspace{4cm}
\caption{
Feynman diagrams for the pair production of the doubly charged
higgsinos in electron-positron collisions.}

\end{figure}

The NLC will, besides the usual
$e^+e^-$ option, be able to work also in $e^-e^-$, $e^-\gamma$ and
$\gamma\gamma$  collision modes.
In these collisions the doubly charged higgsinos $\tilde\Delta^{\pm\pm}$
can be  produced through the following processes:
\be
e^+e^-\to \tilde\Delta^{++}\tilde\Delta^{--},\label{reaction1} \ee
\be e^-e^-\to \tilde\chi^0\tilde\Delta^{--},\label{reaction2} \ee \be
\gamma e^-\to \tilde\l^+\tilde\Delta^{--},\label{reaction3} \ee \be
\gamma\gamma \to \tilde\Delta^{++}\tilde\Delta^{--}.\label{reaction4}
\ee
We have chosen these reactions for investigation because they
all have a clean experimental signature: a few hard leptons and
missing energy. Futhermore, they all have very small background from
other processes.  The fact that    $\tilde\Delta^{\pm\pm}$ carries
two units of electric charge and two units of lepton number and that
it does not couple to quarks  makes the processes (\ref{reaction1}) -
(\ref{reaction4}) most suitable and distinctive tests of the susy
left-right model.

\bigskip

\noindent{\bf  Reaction $e^+e^-\to
\tilde\Delta^{++}\tilde\Delta^{--}$}

\medskip

The triplet higgsino pair production in $e^+e^-$ collision occurs
 through the diagrams presented in Fig. 2.
\begin{figure}
\vspace{6cm}


\caption{
Total cross section for the reaction \(e^+e^-\rightarrow
\tilde{\Delta}^{++}\tilde{\Delta}^{--}\) as a function of the higgsino mass
\(M_{\tilde{\Delta}}\) for two values of selectron mass \(m_{\tilde{l}}\)
at the collision energy 1TeV.}
\end{figure}
In contrast with the  triplet Higgs fields whose  mass
 is in the  TeV scale \cite{Gunion}, the mass of the triplet
higgsino
is given by the susy mass parameter $\mu_2,$
 which is a free parameter. As we mentioned
before, for the reason of naturality its value should not differ too
much from the electroweak breaking scale, i.e.  $\mu_2=O(10^2\,{\rm
GeV})$.

In Fig. 3  the total cross section for the process (\ref{reaction1})
is  presented as a function of the mass of $\tilde\Delta^{--}$ for
the collision energy of $\sqrt{s}=1$ TeV  and for two values of the
selectron mass, $m_{\tilde l}=$ 200 GeV and  400 GeV. As can be seen,
the cross section  for these parameter values is about 0.5 pb and it
is quite constant up to  the threshold region. To have an estimate
for the event rate, one has to multiply the cross section with the
branching ratio of the decay channel of the produced  higgsinos  used
for the search. As pointed out earlier, the favoured decay channel
may be
 \be \tilde\Delta^{--}\to\tilde l^-l^-\to l^-l^-\tilde\chi^0.
\label{llchannel}\ee
Here $l$ can be any of $e,\ \mu$ and $\tau$ with
practically equal probabilities.
The signature of the pair production reaction
(\ref{reaction1}) would be the purely leptonic final state associated
with  missing energy.
In the SM a final state consisting of four charged
leptons  and missing energy can result  from cascade decays. In the
susy left-right  model there are, however, some unique final states
not possible in the SM, namely those with non-vanishing
separate lepton numbers.

\bigskip

\noindent{\bf Reaction $e^-e^-\to \tilde\chi^0\tilde\Delta^{--}$}

\medskip

\begin{figure}
\vspace{4cm}
\caption{Feynman diagrams for the  production of the doubly charged
higgsino in electron-electron collisions.}

\end{figure}
\begin{figure}
\vspace{6cm}


\caption{
Total cross section for the reaction \(e^-e^-\rightarrow
\tilde{\chi}^{0}\tilde{\Delta}^{--}\) as a function of the higgsino mass
\(m_{\tilde{\Delta}}\) for two values of selectron mass \(m_{\tilde{l}}\)
at the collision energy 1TeV.}
\end{figure}
The production of the  triplet higgsino $\tilde\Delta^{--}$ in
electron--electron collision occurs via a selectron exchange in
t-channel  (see Fig. 4).
In Fig. 5
the cross section is  presented as a function of
$M_{\tilde\Delta^{--}}$ for two values of the selectron mass,
$m_{\tilde e}= $ 200 GeV and  500 GeV, at the collision energy
$\sqrt{s}= 1 $ TeV. It is taken into account  that the
final state neutralino mass is related to the triplet higgsino mass
as they both depend on the parameter $\mu_2$. The signature of the
reaction is a same-sign lepton pair created in the cascade decay
(\ref{llchannel}) of $\tilde\Delta^{--}$, associated with the
invisible energy carried by neutralinos. The
two leptons need not be of the same flavour since the $|\Delta L|=2$
Yukawa couplings are not necessarily diagonal. This may be useful for
distinguishing the process from the selectron pair production
$e^-e^-\to\tilde e^-\tilde e^-\to e^-e^- +{\rm neutralinos}$, which
is the leading process for the selectron production in the  susy
version of the Standard  Model. In the SM the final
states $e^-\mu^-$, $e^-\tau^-$ and $\mu^-\tau^-$ are forbidden.

\bigskip

\noindent{\bf Reaction $\gamma e^-\to \tilde\l^+\tilde\Delta^{--}$}

\medskip

The mechanism of producing high-energy photon beams by
back-scattering high intensity laser pulses off the high energy electron
beams  proposed by Ginzburg et al.\cite{photon} allows the NLC to operate
also as a photon collider.
 There are three Feynman diagrams contributing to the photoproduction
reaction  (\ref{reaction3}):  electron exchange in s-channel,
selectron exchange in t-channel and triplet higgsino exchange in
t-channel (see Fig. 6).
\begin{figure}
\vspace{4cm}
\caption{Feynman diagrams for the  photoproduction of the doubly charged
higgsino.}
\end{figure}
\begin{figure}
\vspace{6cm}


\caption{Total cross section for the reaction \(\gamma e^-\rightarrow
\tilde{l}^{+}\tilde{\Delta}^{--}\) as a function of the higgsino mass
\(m_{\tilde{\Delta}}\) for two values of selectron mass \(m_{\tilde{l}}\)
at the electron electron (positron) collision energy 1TeV.}
\end{figure}
In Fig. 7 the total cross section  is
presented as a function of the triplet higgsino mass for the
electron-electron center of mass energy $\sqrt {s_{ee}}=$ 1 TeV.
The cross section is determined by convoluting the photon energy
distribution $P(y)$, i.e.  $\sigma(s_{ee})=\int dyP(y)\sigma(s_{e\gamma})$.

The experimental signature of the reaction is three lepton final
state  associated with missing energy. The positive lepton is any
lepton, and the two negative ones can be any combination of the
electron, muon and tau, provided the triplet higgsino coupling is not
diagonal. A suitable choice of the final state will cut down the
SM background coming e.g. from the reaction  $e^-\gamma
\to e^-Z^*$.  The cross section is above  {\cal O}(100 fm) for a
large range of the masses $M_{\tilde\Delta^{--}}$ and $m_{\tilde e}$,
providing hence a good potential for the discovery of
$\tilde\Delta^{--}$.

\bigskip

 \noindent{\bf Reaction $\gamma\gamma \to
\tilde\Delta^{++}\tilde\Delta^{--}$}

\medskip

This reaction is an alternative of, but not  competitative with, the
reaction (\ref{reaction1}) for producing a doubly charged  higgsino
pair. Feynman diagram of the process is presented in Fig. 8.
\begin{figure}
\vspace{4cm}
\caption{Feynman diagrams for the  production of the doubly charged
higgsinos in photon-photon collisions.}

\end{figure}
Because
the photon energies are not monochromatic but broadly distributed, no
sharp threshold will be visible  in the production cross section.
Moreover, the maximum collision energy will be some 20\% less than
the $e^+e^-$ energy. On the other hand, the only unknown parameter in
the process is the mass $M_{\tilde\Delta^{--}}$ as the couplings are
completely determined by the known electric charge of the higgsino.
\begin{figure}
\vspace{6cm}


\caption{
Total cross section for the reaction \(\gamma\gamma\rightarrow
\tilde{\Delta}^{++}\tilde{\Delta}^{--}\) as a function of the higgsino mass
\(m_{\tilde{\Delta}}\)
at the electron electron collision energy 1TeV.}

\end{figure}
 The cross section of the reaction as a function of
$M_{\tilde\Delta^{--}}$  is given in Fig. 9 for the collision energy
$\sqrt{s_{ee}}$= 1 TeV. The experimental signature of the reaction
will be of course the same as for the process (\ref{reaction1}), i.e.
four charged leptons associated with missing energy. The cross
section is large because of the photon coupling to electric charge.

\bigskip

\noindent {\bf Slepton pair production }

\medskip

If the doubly charged higgsinos turn out to be too heavy to be
produced in the NLC one can study their possible contributions as the
intermediate virtual particles to the production of lighter  susy particles.
One such process is the slepton pair production (see Fig. 10)

\be e^+e^-\to\tilde l^+\tilde {l}'^-, \label{reaction}\ee

\noindent  where $\tilde l,\: \tilde l'= \tilde e,\tilde\mu,
\tilde\tau$.
Experimentally, the slepton pair production is possibly one of the
first susy processes to be seen since
sleptons are supposed to be relatively light among the
superpartners of the ordinary particles.
Compared with the MSSM there is one extra
$s$-channel diagram  involving the heavy neutral gauge boson $Z_2,$
five extra $t$-channel diagrams due to the larger number of neutralinos
and one new $u$-channel diagram in our model.

In Fig. 11 we present the total cross section of the selectron pair
\begin{figure}
\vspace{4cm}
\caption{Feynman diagrams for the slepton pair production in
electron positron collisions.}
\end{figure}
\begin{figure}

\vspace{6cm}


\caption{ The total cross section
$\protect \sigma(e^+e^-\to \tilde e^+_L\tilde e^-_L)+\sigma(e^+e^-\to
\tilde e^+_R\tilde e^-_R)+ 2\sigma(e^+e^-\to \tilde e^+_L\tilde
e^-_R)$ as a function of the selectron mass $\protect m_{\tilde l}$ (a) for
the collision energy $\protect \sqrt{s}=200$ GeV and triplet higgsino mass
$\protect M_{\tilde\Delta}= 110$ GeV, (b) for $\protect \sqrt{s}=1$ TeV,
$\protect M_{\tilde\Delta}= 300$ GeV. LRM I (II) refer to  two
supersymmetric left-right models and MSSM I (II) to two versions
of the minimal supersymmetric Standard Model described in the text.}
\end{figure}
production as a function of the selectron mass $m_{\tilde e}$
for a fixed triplet higgsino
mass $M_{\tilde\Delta}$. Fig. 11a corresponds to the
situation at LEP200 with $\sqrt{s}=200$ GeV (here
$M_{\tilde\Delta}= 110$ GeV) and Fig. 11b at a linear collider
with  $\sqrt{s}=1$ TeV ($M_{\tilde\Delta}= 300$ GeV).

The cross sections in the two \lr cases differ slightly
because in LRM I the $t$-channel
processes are supressed since the higgsino couplings are small, in
LRM II there is no such suppression.
For comparison we have plotted in Fig. 11 also the corresponding
cross section in  the MSSM I (II).  As one can see, the  cross sections
in the susy left-right model are systematically appreciably larger than
in the MSSM.  This is due to two factors, firstly the number of
gauginos is larger and secondly the triplet higgsino contribution
is large, though dependent on the unknown triplet higgsino
coupling to the electron and selectron.

The most intriguing difference between the susy
left-right model  and the MSSM with respect to the
slepton pair production is the existence of  the $u$-channel
process of Fig. 10c. This reaction
occurs only for a right-handed electron and a
left-handed positron, whereas in the $s$- and $t$-channel
processes all chirality combinations may enter. Use of polarized
beams  could therefore give us more information of the triplet
higgsino contribution. Assuming that the decay mode $\tilde e\to
e\tilde\chi^0_1$ is dominant,  we present in Fig. 12 the angular
distribution of the final state electron $e^-$ in
the case the electron is right-handedly and  positron
left-handedly polarized ($P_{+-}$) and in the opposite case
($P_{-+}$)
for $\sqrt{s}= 1$
TeV, $m_{\tilde e}=M_{\tilde\Delta}= $ 300 GeV.
In model LRM I (Fig. 12a)  the $t$-channel contributions are
suppressed since the light neutralinos are mainly consisting of
the  higgsinos.  The  distribution $P_{+-}$ is larger and it is
slightly peaked in the backward direction because of the
$u$-channel contribution. In  model LRM II (Fig. 12b)
both $t$-channel (forward peak) and $u$-channel (backward peak)
 contributions are observable. The latter is  absent in the MSSM,
of course.

In the MSSM with the unification assumption the right-selectron is
lighter than the left-selectron \cite{heavyeL}. If only $\tilde
e_R$'s are produced, the difference between the MSSM and the
supersymmetric left-right model would be especially large
and observable in the NLC.
\begin{figure}
\vspace{6cm}


\caption{ The angular distribution of the final
state electron in the cascade process $\protect e^+e^-\to \tilde e^+\tilde
e^-\to e^+e^-\tilde\chi^0_1\tilde\chi^0_1$ for $\protect \sqrt s=1$  TeV,
$\protect M_{\tilde\Delta}= m_{\tilde e}=300$ GeV in the model (a) LRM I,
(b) LRM II. $\protect P_{+-}$ corresponds to the case where the incoming
electron has positive and the incoming positron has negative
longitudinal polarization, and $\protect P_{-+}$ corresponds to the
opposite case.
}
\end{figure}

\section{Conclusions}

Phenomenologically the most intriguing prediction of the
supersymmetric \ssu model is the existence of doubly charged
higgsino $\tilde\Delta^{++}$.
It carries two units of lepton number and it has a clear decay signature:
two same sign leptons, which are not necessarily the same type, and
the missing energy.
The production cross sections of $\tilde\Delta^{++}$ in $e^+e^-$, $e^-e^-$,
$e^-\gamma$ and $\gamma\gamma$
collision modes of the NLC are at the level of pb.
Its contribution to the slepton pair production processes should be
observable in angular distributions of the process.

\vspace{1cm} \large \noindent Acknowledgements \normalsize
\vspace{0.5cm}

 The work has been supported by the Finnish Academy of Science. One
of the authors (M.R.) expresses his gratitude to the Viro-s\"a\"ati\"o,
Emil Aaltosen S\"a\"ati\"o and Wihurin Rahasto
foundations for  grants.

\end{document}